\documentclass[prl,twocolumn,superscriptaddress]{revtex4}

\begin{document}

\title{Comment on \textquotedblleft Mass and Width of the Lowest Resonance in QCD\textquotedblright}

\author{Frieder Kleefeld}
\email{kleefeld@cfif.ist.utl.pt}
\homepage{http://cfif.ist.utl.pt/~kleefeld/}
\affiliation{Doppler Institute for Mathematical Physics
and Applied Mathematics \& Nuclear Physics Institute (Department of Theoretical
Physics), Academy of Sciences of  Czech Republic, 250 68 \v{R}e\v{z} near Prague, Czech Republic}
\affiliation{Centro de F\'{\i}sica das Interac\c{c}\~{o}es Fundamentais (CFIF),
Instituto Superior T\'{e}cnico,
Av. Rovisco Pais,
P-1049-001 LISBOA,
Portugal}

\date{\today}

\begin{abstract}
I. Caprini's, G. Colangelo's, and H. Leutwyler's (CCL) article ``Mass and Width of the Lowest Resonance in QCD'', Phys. Rev. Lett. 96, 132001 (2006) [arXiv:hep-ph/0512364], is critically reviewed. The present comment is devoted to complement a recent experimental discussion (D.V. Bugg, J. Phys. G 34, 151 (2007) [arXiv:hep-ph/0608081]) of short-comings in the CCL analysis, by presenting theoretical arguments pointing at a serious flaw in the theoretical formalism used by CCL, and also at the unlikeliness of their tiny error bars in the $\sigma$-meson mass and width. The criticism made in the comment applies analogously to the analysis on the $\kappa$-meson mass performed in the article ``The K0*(800) scalar resonance from Roy-Steiner representations of pi K scattering'' published as S. Descotes-Genon and B. Moussallam, Eur. Phys. J. C 48, 553 (2006) [arXiv:hep-ph/0607133].
\end{abstract}

{\bf\noindent Comment on \textquotedblleft Mass and Width of the Lowest Resonance in QCD\textquotedblright} \\[3mm]
In a recent Letter \cite{Caprini:2005zr}, in which \textit{no} use is made of QCD, I.~Caprini, G.~Colangelo, and H.~Leutwyler (CCL) repeated an unmentioned analysis of $\pi\pi$ scattering from 1973  \cite{Pennington:1973xv}, based on the Roy equations (REs), to make out a case for the existence of a scalar $I=0$ resonance $f_0(441)$, listed in the PDG tables \cite{Yao:2006px} as $f_0(600)$ and known as $\sigma$-meson. The primary aspect resulting of the CCL analysis is the claimed {\em model-} and {\em parametrization-independent} determination of a $\sigma$-pole mass of $(441^{+16}_{-8}-\frac{1}{2}\,i\,544^{+18}_{-25})$~MeV implying unprecedented small error bars. Moreover, the latter result is incompatible with very recent experimental findings, i.e., $(500\pm 30-\,i\,(264\pm 30))$~MeV~\cite{Bugg:2006gc} and $(541\pm 39-\,i\,(252\pm 42))$~MeV~\cite{Ablikim:2004qn,Bugg:2005xz}, as well as with a combined theoretical analysis yielding $((476$--$628)-\,i\,(226$--$346))$~MeV \cite{vanBeveren:2006ua}. The present comment will be devoted to complement a recent experimental discussion \cite{Bugg:2006gc} of short-comings in the CCL analysis, by presenting theoretical arguments pointing at a serious flaw in the theoretical formalism used by CCL, and also at the unlikeliness of their tiny error bars in the $\sigma$ mass and width. 
The simplest way to identify this flaw in Ref.\ \cite{Caprini:2005zr} also present in the corresponding results \cite{Descotes-Genon:2006uk} of S.~Descotes-Genon and B.~Moussallam (DM) on the scalar meson $K^\ast_0(800)$ in the context of Roy-Steiner equations (RSEs), is to recall a warning statement  by G.F.~Chew and S.~Mandelstam (CM) from 1960 (see footnote 6 of Ref.~\cite{Chew:1960iv}). CM state that if a strongly interacting particle with the same quantum numbers as a pair of pions should be found, then corresponding poles must be added to the double-dispersion representation, whether or not the new particle is interpreted as a two-pion bound state. It should be emphasized that this statement does not only apply to possible bound-state (BS) poles of the S- or T-matrix in the physical sheet (PS) of the complex $s$-plane, but also to any kind of virtual BS poles and resonance poles in the unphysical sheet (US). This is justified from first principles by reviewing briefly how dispersion relations (DRs) are to be derived on the basis of Cauchy's integral formula $t(s)=(2\pi i)^{-1}\oint dz \; t(z)/(z-s)$ which holds for a function $t(s)$ \textit{analytic} in the domain encirculated by the closed integration contour. As the so-called ``matching point'' of CCL (and DM) is located \textit{in the US}, the closed integration contour yielding the REs/RSEs must extend \textit{also to the US} where the S- and T-matrix poles for scalar isoscalar $\pi\pi$-scattering are found. Excluding these poles situated at $s_j$ ($j=1,\ldots, n$) from the integration contour and assuming $t(s\rightarrow \infty)\rightarrow 0$ sufficiently fast one obtains the well known (here) unsubtracted DRs $t(s)= \sum_{j=1}^n r_j/(s-s_j) - \frac{1}{\pi} \int_{L,R} dz \,\mbox{Im}[t(z)]/(s-z+i\varepsilon)$,
where L/R denotes the left-/right-hand cut, and $r_j$ is the residue of $t(s)$ at the corresponding pole $s_j$. According to CCL, REs/RSEs are twice-subtracted DRs yielding 
\begin{eqnarray} t(s) & = & t(s_0) + (s-s_0) \; t^\prime(s_0) + \sum\limits_{j=1}^n \frac{(s_0-s)^2 \,r_j}{(s_0-s_j)^2(s-s_j)} \nonumber \\ & - & \frac{1}{\pi} \int_{L,R} dz\;\frac{(s_0-s)^2 \; \mbox{Im}[t(z)]}{(s_0-z+i\varepsilon)^2(s-z+i\varepsilon)} \; , \label{eq1}
\end{eqnarray}
where the subtraction point $s_0$ used by CCL appears to be the $\pi\pi$ threshold, as CCL perform the identification $t(s_0)=a_0^0$ and $t^\prime(s_0)=(2a_0^0-5a^2_0)/(12 m^2_\pi)$ with $a^I_0$ being S-wave scattering lengths for isospin $I=0,2$. It is now easy to see that the REs/RSEs considered by CCL and DM {\em disregard the pole terms} (PTs) in the DRs (yielding $r_j = 0$), despite the presence of poles in the US that are claimed to exist by observing respective S-matrix zeros in the PS. As the $s$-dependence of the disregarded $\sigma$- and $f_0(980)$-PTs in the vicinity of the $\pi\pi$- and $KK$-theshold is clearly {\em non-linear}, it is to be expected on grounds of dispersion theory that the S-matrix poles predicted by CCL will {\em not} coincide with the actual ones to be determined yet by CCL for self-consistency reasons. An analogous statement applies to the results of DM. Moreover will the inclusion of PTs in REs/RSEs not only reinstate dispersion theoretic self-consistency, yet also yield a {\em significant} change in the resulting $\sigma$- and $K^\ast_0(800)$-pole positions, which unfortunately will enter now via the PTs as unknown parameters the REs/RSEs to be solved. Hence the inclusion of PTs in REs/RSEs will yield an uncertainty of pole positions which is likely to be of the order of the one estimated in Ref.\  \cite{Bugg:2006gc} and therefore much larger than the error bars presently claimed by CCL and DM being even without taking into account PTs for at least two reasons clearly {\em parametrization-dependent}: (1)~the extrapolation of the two particle phase space to the complex $s$-plane and below threshold invoked by CCL and DM is known to be speculative and even unphysical as it yields e.g.\ in the approach of DM scattering below the pseudo-threshold; (2)~standard chiral perturbation theory (ChPT) disregarding (yet) non-perturbative PTs relates claimed values for scalar scattering lengths and their (too) tiny error bars entering REs/RSEs lacking (yet) PTs to scalar square radii $\left<r_S^2\right>$ the presently used (too) high values of which yield chiral symmetry breaking (ChSB) of the order of 6-8\% being much larger than $3$\% as observed in Nature. A revision of the analysis of CCL and DM by taking into account PTs in REs/RSEs and ChPT would be highly desirable to reconcile their results with Refs.~\cite{Bugg:2006gc}-\cite{vanBeveren:2006ua} and to improve the poor description of the resonance $K^\ast (892)$ in the approach of DM. \\[2mm]
\noindent Frieder Kleefeld ${}^{1,2,3}$ \\[1mm]
${}^{1}$ Present address: Pfisterstr.\ 31, 90762 F\"urth, Germany \\[1mm]
${}^{2}$ Doppler \& Nucl.~Physics Institute (Dep.~Theor.~Phys.), Academy of Sciences of  Czech Republic; collaborator of the CFIF,
Instituto Superior T\'{e}cnico,
LISBOA,
Portugal\\[1mm]
${}^{3}$ Electronic address: {\sf kleefeld@cfif.ist.utl.pt}

This work has been supported by the FCT of the \textit{Minist\'{e}rio da Ci\^{e}ncia, Tecnologia e Ensino Superior} of Portugal,
under contracts POCI/FP/63437/2005, PDCT/FP/63907/2005 and the Czech project LC06002.\\[1mm]
Conversations with E.\ van Beveren, D.~V.\ Bugg, \mbox{J.~Fischer}, A.\ Moussallam, G.\ Rupp, M.~D.\ Scadron and M.\ Znojil are gratefully acknowledged. 

\end{document}